\documentclass[pre,a4paper,superscriptaddress,twocolumn,showpacs,amsmath,amssymb,floatfix]{revtex4}
\pdfoutput=1
\usepackage{graphicx}
\usepackage{amssymb}
\usepackage{float}

\def\eps{\varepsilon}

\usepackage{color}

\begin{document}

\title{Goal quest for an intelligent surfer
moving in a chaotic flow}

\author{Klaus M.~Frahm}
\affiliation{\mbox{Laboratoire de Physique Th\'eorique, IRSAMC, 
Universit\'e de Toulouse, CNRS, UPS, 31062 Toulouse, France}}
\author{Dima L.~Shepelyansky}
\affiliation{\mbox{Laboratoire de Physique Th\'eorique, IRSAMC, 
Universit\'e de Toulouse, CNRS, UPS, 31062 Toulouse, France}}

\date{June 26, 2023}

\begin{abstract}
   We  consider a model of an intelligent surfer moving on 
  the Ulam network generated by a chaotic dynamics in the Chirikov standard map. 
  This directed network is obtained  by the Ulam method with
  a division of the phase space in cells of fixed size 
  forming the nodes of a Markov chain.
  The goal quest for this surfer is to determine the network path 
  from an initial node A to a final node B with minimal resistance 
  given by the sum of inverse transition probabilities. 
  We  develop an  algorithm
  for the intelligent surfer that allows to perform the quest in a small number
  of transitions which grows only logarithmically with the network size.
  The optimal path search is done on a fractal intersection set formed by 
  nodes with small Erd\"os numbers of the forward and inverted networks. 
  The intelligent surfer exponentially outperforms
  a naive surfer who tries to minimize its phase space distance to the target B.
  We argue that such an algorithm provides new hints for motion control in chaotic flows.
\end{abstract}



%

\maketitle

\section{Introduction} 
\label{sec1}

The  time evolution of probability
of chaotic map dynamics in a continuous phase
space is described by the Perron-Frobenius operator
(see e.g. \cite{lichtenberg,cvitanovic}).
In 1960 Stanislaw Ulam proposed a method \cite{ulam}
which gives a  discrete finite cell 
approximate of the Perron-Frobenius operator for a
completely chaotic map. In this method, known as the Ulam method, 
the transition probabilities from one cell to others
are obtained via an ensemble of trajectories
generating the probabilities of Markov transitions
between cells in one map iteration. This gives a
finite size Markov matrix of transitions on the corresponding
Ulam network. For one-dimensional (1D)
completely chaotic maps the convergence of the discrete
cell description of this Ulam approximate of the 
Perron-Frobenius operator (UPFO)
to the continuous chaotic dynamics
has been rigorously proven in \cite{li}.
 The properties of UPFO were analysed in
\cite{tel,kaufmann,froyland2007} and
\cite{ding,liverani,froyland2008a,froyland2008b}
respectively for 1D and 2D chaotic maps.

The  finite cell size of UPFO effectively introduces
a finite noise in dynamical equations with an
amplitudeproportional to a  cell size. For generic
symplectic maps with divided phase space,
containing chaotic components and integrable islands
like the Chirikov standard map \cite{chirikov},
such a noise destroys invariant Kolmogorov-Arnold-Moser (KAM) curves
\cite{lichtenberg,chirikov}
and the  original Ulam method does not provide a correct description
of dynamics in such cases. In \cite{frahmulam,ulampoincare}
it was shown that a generalized Ulam method
resolves the above problem and provides a correct transition description 
for the chaotic component bounded by original KAM curves.
In this generalized method the  Markov transitions are obtained  
with specific trajectories starting only inside one chaotic component
thus generating Markov transitions only between cells of
the same chaotic component. For such a case it was established 
\cite{frahmulam,ulampoincare} 
that the spectrum of the finite size UPFO matrix
converges to a limiting density as the cell size approaches zero.

In \cite{smulam} it was shown that the Ulam networks generated
by the generalized Ulam method have the properties of small-world networks 
similar to the {\it six degrees of separation}
which appear in social networks of people \cite{milgram},
actors, power grids, biological and 
other networks \cite{strogatz,newman,dorogovtsev}.
Thus, as demonstrated in \cite{smulam}, for the Ulam networks of
symplectic maps the number of degrees of separation, or the Erd\"os number \cite{erdos},
grows only logarithmically with the network size for the regime of strong chaos.
This growth is related to the Kolmogorov-Sinai entropy and the positive
Lyapunov exponent of chaotic dynamics \cite{lichtenberg,cvitanovic,chirikov}
which leads to an exponential divergence
of initially nearby trajectories. Thus even in Ulam networks of huge
size the Erd\"os number remains rather moderate and any
cell can be reached in about ten or several tens of transitions
by a random surfer.

The concept of random surfer is broadly used for
the construction of the Google matrix $G$ of various directed networks
and is at the foundations of the Google search engine \cite{brin,meyerbook,rmp2015}.
Such a surfer jumps randomly following cell links to other cells according 
to the probabilities of Markov transitions. 
Due to a damping factor of the Google matrix
and in a case of dangling nodes (or cells) 
a jump can go to any other cell but the related transition 
probabilities are very small in comparison to the probabilities 
of direct links. 
In the case of Ulam networks of symplectic maps,
e.g. the Chirikov standard map,
there are no dangling nodes neither damping factor and 
there is only a relatively small number (about 10) of 
transitions from one cell to other cells \cite{frahmulam,ulampoincare,smulam}.

Thus for the Ulam networks of chaotic symplectic maps
it is interesting to consider not only a random surfer but also to analyze a behavior of
an intelligent surfer who has a goal quest 
starting from an initial node $A$
to reach another node $B$ following a path with
highest probability of jumps and a minimal number of jumps.
So, in this work we analyze the optimal strategies
and algorithms to be followed by an intelligent
surfer resolving the goal quest
and moving on such Ulam networks.

We note that the problem of shortest path detection 
has been studied actively in computer science
for various types of networks
(see e.g. \cite{path1,path2,path3,path4}
and Refs. therein). However, here we have 
a rather specific case of the Ulam networks
obtained from a chaotic symplectic dynamics
and we think that our analysis provides
a new useful view on the properties of chaos. 
Also, we consider a somewhat different problem where not the 
path length as such but a different quantity, being the {\em resistance}, is minimized.

The paper is organized as follows:
Sections \ref{sec2} and \ref{sec3} remind and present the main properties of the
Chirikov standard map, the construction of the corresponding 
Ulam network and also the definition of the {\em resistance} 
of a given network trajectory. 
Section \ref{sec4} describes an efficient algorithm for an intelligent surfer 
to determine optimal 
trajectories with minimal resistance between two given nodes using 
a specific intersection set $S_I$ of nodes with small Erd\"os and inverse Erd\"os numbers. 
In Section \ref{sec5}, numerical results based on this algorithm are presented, 
in Section \ref{sec6} the fractal structure of the set $S_I$ is studied and 
in Section \ref{sec7} the dependence of the minimal resistance on the network size 
is discussed. In Sections \ref{sec8} and \ref{sec9}, alternative models 
of a naive surfer or an intelligent surfer with limited resources are introduced 
and studied and the discussion of the results is presented in Section \ref{sec10}.
Additional data and Figures are presented in the Supplementary Material (SupMat).

\section{Chirikov standard map}
\label{sec2}

We consider the Ulam network for the Chirikov standard map \cite{chirikov}.
This map captures the important generic features of chaotic
Hamiltonian dynamics and 
finds a variety of applications for the description of real physical systems
(see e.g. \cite{stmapscholar}). The map has the form:
\begin{equation}
\label{eq_stmap}
{\bar p} = p + \frac{K}{2\pi} \sin (2\pi x) \; , \;\; 
{\bar x} = x + {\bar p} \;\; ({\rm mod} \; 1) \;.
\end{equation}
Here bars mark the variables after one map iteration,
$p,x$ are conjugated variables of momentum and coordinate and 
$K$ is the usual chaos parameter. 
We consider the dynamics to be periodic on  a torus such that
$0 \leq x \leq 1$, $0 \leq p \leq 1$. It has been argued that the last
KAM curve, with the golden rotation number, is destroyed 
at critical $K_c=K_g=0.971635406...$ \cite{mackay}.
A more rigorous mathematical analysis \cite{percival}
proved that all KAM curves are destroyed
for $K \ge 63/64$ while numerical computations 
showed that  $K_c-K_g  < 2.5 \times 10^{-4}$  \cite{chirikov2000}.
Thus, the golden KAM curve at  $K_c=K_g$ looks to be
the last one to be destroyed (see also the review \cite{meiss}).
For values $K\ge 2.5$ considered in this work, the dynamics is 
clearly globally chaotic but a certain fraction of stable non-chaotic 
islands is possible. The construction of the Ulam network, explained 
in the next section, automatically avoids such regions. 

\section{Construction of Ulam network}
\label{sec3}

The Ulam network and related UPFO for the map (\ref{eq_stmap}) 
are constructed as described in \cite{frahmulam}. For pedagogical reasons 
we give a brief summary here. 
First, exploiting the symmetry $x\to 1-x$ 
and $p\to 1-p$ the phase space is reduced to the region $0\le x<1$ 
and $0\le p< 0.5$ and then it is divided into $M\times (M/2)$ cells 
with certain integer values $M$ in the range $200\le M\le 3200$.
Here we study three values of the chaos parameter being $K=2.5,\,5,\,7$. 
We consider 
one very long trajectory of $10^{12}$ iterations with initial condition 
at $x=p=0.1/(2\pi)$, which is in the chaotic component, and 
count the number of transitions $N_{i\to j}$ from a cell $i$ to a cell $j$. 
This allows to determine the classical transition probabilities 
$p_{i\to j}=N_{i\to j}/\sum_j N_{i\to j}$. The index number $i$ associated 
to each cell is constructed at the same time, i.e. each time the trajectory 
enters a new cell, not yet visited before, the value of $i$ is increased 
by one and attributed to this new cell. 
Depending on the value of $K$ it is possible that 
there are stable islands or other non-accessible regions where the 
trajectory  never enters and therefore the network size $N$ is typically 
$N<M^2/2$ but for the considered $K$ values with $K\ge 2.5$ the 
fraction of stable islands is quite modest such that $N\approx M^2/2$. 
The non visited cells due to such islands do not contribute to the Ulam network. 

In practice, we perform trajectory iterations only for the 
largest two values $M=3200$, $M=2240$ (the latter not used in this work) and 
apply an exact renormalization 
scheme to reduce successively the value of $M$ by a factor of 2 to 
smaller values of $M$. In this work, we limit ourselves to the cases 
$M=200,400,800,1600,3200$. 

We remind that the original Ulam method \cite{ulam} 
computes the transition probabilities from one cell 
to other cells with many random initial conditions per cell but for the 
Chirikov standard map this procedure is less convenient since it 
causes an implicit coarse graining with diffusion into the stable islands 
or other classically non-accessible regions which we want to avoid. 

The matrix $G_{ji}=p_{i\to j}$ corresponds to an Ulam approximate
of Perron-Frobenius  operator (called UPFO)
satisfying $G_{ji}\ge 0$ and the column 
sum normalization $\sum_j G_{ji}=1$. In \cite{frahmulam,ulampoincare}, the 
(complex and real) eigenvalues $\lambda_j$ with $|\lambda_j|\le 1$ 
and eigenvectors of $G$ were analyzed and studied in detail. In absence of 
absorption, which is the case in this work, the leading eigenvalue 
is $\lambda_1=1$ and its eigenvector components give the global ergodic 
rather constant density of cells which are in the chaotic region 
of the phase space accessible from the initial trajectory.
Due to ergodicity the choice of the initial trajectory is not important
if it starts inside the global chaotic component.

In \cite{smulam}, among other things, also the distribution of 
the outgoing link number $N_l$ of a given cell $i$, 
i.e. number of non-vanishing matrix elements $G_{ji}>0$ for fixed $i$, 
were analyzed. In particular, it was 
found \cite{smulam} that the maximal possible value of $N_l$ is 
$N_l^{(\rm max)}=2\lceil 2+K\rceil$ and statistically mostly even 
values of $N_l$ are possible (rare odd values are possible 
due to boundary effects and mostly for small $M$).

For a given trajectory $i_j\in\{1,\ldots,N\}$, 
$j=0,\ldots,l$ between 
two cells $A=i_0$ and $B=i_l$, we define the {\em resistance} $R$ as~:
\begin{equation}
\label{resdef}
R=\sum_{j=1}^{l} \frac{1}{p_{i_{j-1}\to i_{j}}}
\end{equation}
representing the sum of inverse transition probabilities over each 
leg of the trajectory. Of course for each transition $i_{j-1}\to i_{j}$, 
we require $p_{i_{j-1}\to i_{j}}>0$, i.e. the network allows 
this transition with an existing link.
According to (\ref{resdef}) a trajectory, from an 
initial cell $A$ to a finite cell $B$, with a minimal
resistance value $R$ follows a path with
maximal probabilities along the path and small value $l$ 
of the path length. In the following, we study the properties of
such optimal trajectories with minimal $R$ values.

\section{Algorithm to determine optimal trajectories} 
\label{sec4}

In the following, we consider the Erd\"os number $N_{E,A}(C)$ 
(or number of degrees of separation) 
\cite{erdos,dorogovtsev} representing the minimal number of links 
necessary to reach indirectly the specific node $C$  via other intermediate 
nodes from a particular node $A$, also called the hub. 
In this context, only the existence of a link between two nodes is important 
and the precise value of  transition chain of probabilities is not relevant, 
as long as it is non-vanishing. 
For a given hub $A$ one can quite efficiently compute $N_{E,A}(C)$ for 
all other nodes $C$ of the network in a recursive way. For this, 
one considers  
at a certain level $n\ge 0$ the set of nodes $C$ with $N_{E,A}(C)=n$ 
(at initial level $n=0$ this set only contains the hub $A$ itself) and 
one determines the corresponding set at level $n+1$ as the set of all 
nodes (i) which can be reached by a direct link from the level $n$ set and 
(ii) which are not member of a former set of level $\le n$ 
(which can be rapidly verified by keeping a list of all used nodes up 
this level). Initially, the size of these sets grows exponentially 
with $\sim N_l^n$ elements where $N_l$ is the typical or average number 
of outgoing links per node (with $N_l\le 2\lceil 2+K\rceil$ for Ulam networks 
of the Chirikov map \cite{smulam}) but 
the exponential growth is quickly saturated once the majority of the network 
is occupied and the final complexity of this algorithm is $\sim N_l\,N$ 
since essentially for each node we have to perform $N_l$ tests if the outgoing 
link is going to a new unused node or to an already occupied node. 

Our aim is to determine the optimal trajectory that goes 
from a node $A$ to another node $B$ and which minimizes its 
resistance $R$ defined above in (\ref{resdef}). 
The node $A$ is chosen at position $(x_0,p_0)=(0.1,\,0.1)/(2\pi)$ 
which is also the initial position of the unique and very long classical 
trajectory used to compute the Ulam network \cite{frahmulam}. 
The final node $B$ is chosen at position $(x_0,p_0)=(0.84,\,0.4)$ for $K=5$ 
and $K=7$. However, for $K=2.5$, we choose a slightly different initial 
position for $B$ at $(x_0,p_0)=(0.92,\,0.4)$ in order to assure that $B$ is 
in a zone of modest value of $N_{E,A}(B)\le 12$ which is quite important to 
determine efficiently the optimal trajectories.
A change of cell positions of $A, B$ does not affect the results if $A, B$ are taken inside
a component of strong chaos (so we avoid the vicinity of integrable islands).

The minimal length $l_{\rm min}$ of a trajectory between $A$ and $B$, 
in number of links, is simply per definition the Erd\"os number 
$l_{\rm min}=N_{E,A}(B)$ and there are no 
trajectories between $A$ and $B$ with length $l<l_{\rm min}$. 
Furthermore, for any trajectory $i_j$, $j=0,1,2,\ldots$ with $i_0=A$, we 
know that $N_{E,A}(i_j)\le j$, i.e. the Erd\"os number of the 
intermediate cell $i_j$ after $j$ steps with respect to the 
initial cell $A$ cannot be larger than the number of $j$ steps 
since, simply per definition, the Erd\"os number is the minimal 
number of steps to reach the cell $i_j$. 

For practical reasons it is necessary to limit the search of optimal 
trajectories to some finite maximal length $l_{\rm max}$ and we choose 
as $l_{\rm max}=l_{\rm min}+4=N_{E,A}(B)+4$, i.e. we allow potentially 
for 4 additional legs in the trajectory with respect to the minimal length, 
assuming that the trajectories with minimal resistance values have also 
a length $l$ being close to $l_{\rm min}$. In most cases, except 
for $K=2.5$ and small values of $M$, this is sufficient to find the 
optimal trajectory. 
Such trajectories are therefore limited to nodes $C$ with 
Erd\"os number below $l_{\rm max}$, i.e. $N_{E,A}(C)\le l_{\rm max}$. 
We call the set of such nodes $S_{E,A}$. 

In order to find practically the optimal trajectory one has in 
principle to apply a recursive search starting from the hub but 
this procedure has an exponential complexity $\sim l_{\rm max}^{N_l}$ 
resulting in a large number of test trajectories that cannot reach the 
selected end point $B$. 
To simplify this search we note that for each trajectory $A\to B$ 
in the Ulam network we have also the inverted trajectory $B\to A$ 
in the inverted Ulam network (with all links inverted). Therefore, 
we also compute the {\em inverse Erd\"os} number $N^*_{E,B}(C)$ 
for all nodes $C$ as the Erd\"os number of the inverted Ulam network 
with respect to the final point $B$ as hub and determine 
the set $S^*_{E.B}$ of nodes $C$ with $N^*_{E,B}(C)\le l_{\rm max}$. 
The cells $i_j$, $j=0,\ldots, l$ 
of each possible trajectory $A\to B$ (with length $l\le l_{\rm max}$
and $i_0=A$, $i_l=B$) 
belong to both sets $S_{E,A}$ and $S^*_{E.B}$ and therefore they 
are limited to cells belonging to the intersection 
$S_{E,A} \cap S^*_{E.B}$. Actually, each cell $i_j$ of such a trajectory 
even satisfies a stronger condition for its iteration number $j$:
\begin{equation}
\label{trajcond}
j\le l-N^*_{E,B}(i_j)\le l_{\rm max}-N^*_{E,B}(i_j)
\end{equation}
since the iteration number $j^*=l-j$ of the inverted trajectory 
obviously satisfies $N^*_{E,B}(i_j)\le j^*$ in a similar way as 
$N_{E,A}(i_j)\le j$. From the latter inequality and (\ref{trajcond}), 
we find that $i_j\in S_I$ where $S_I$ is defined as the set of cells $C$
satisfying the condition 
\begin{equation}
\label{trajcond2}
N_{E,A}(C)+N^*_{E,B}(C)\le l_{\rm max}\ .
\end{equation}
Therefore each possible trajectory $A\to B$ of length $l\le l_{\rm max}$ 
is limited to nodes of the set $S_I$. In the following, we also call $S_I$ 
{\em intersection set} because it is a (smaller) subset 
of the intersection $S_{E,A} \cap S^*_{E.B}$ since (\ref{trajcond2}) 
implies both conditions for the subsets $S_{E,A}$ and $S^*_{E.B}$. 
However, we mention that the inverse is not true implying that $S_I$ is a strictly 
smaller subset of $S_{E,A} \cap S^*_{E.B}$ and not equal to the latter.

The recursive search of optimal trajectories can be greatly improved 
by limiting the recursion at each step to possible nodes $i_j$ satisfying 
the (2nd) inequality (\ref{trajcond}). In SupMat Fig.~S1, 
the number $N_T$ of such possible trajectories 
is shown for the parameter choices 
of $A$, $B$, $M$, $K$ and $l_{\rm max}$ given above. 
Even though this value can be quite large (largest value between 
$10^8$ and $10^9$) it is typically quite modest and clearly in the range 
where it is comparable to the cost of computation of both 
sets of Erd\"os numbers (which has to be done in advance; a few seconds 
on a single processor of a modern computer for $M=3200$).

\section{Trajectories with minimal resistance}
\label{sec5}

\begin{figure}[t]
\begin{center}
  \includegraphics[width=0.46\textwidth]{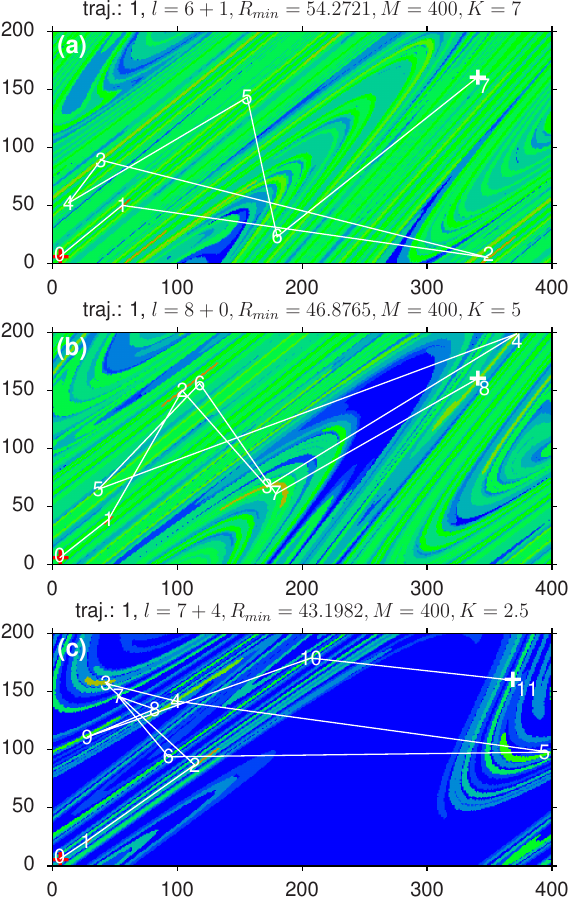}
\end{center}
\caption{\label{fig1}
The white lines show the optimal trajectory between 
the initial point $A$ (red cross) and final point $B$ (white cross) 
with minimal resistance $R$ (see text for the definition). The white numbers 
correspond to  the iteration number $j=0,\ldots, l$ of 
the trajectory points $i_j$ with $l$ being the length of the trajectory. 
The color plot shows the set 
$S_{E,A}$ of cells $C$ with Erd\"os number $N_{E,A}(C)\le 
l_{\rm max}=N_{E,A}(B)+4$ where red/green/light blue/full blue colors correspond
to maximal/intermediate/small/negative (if $C\not\in S_{E,A}$) values of
the difference $l_{\rm max}-N_{E,A}(C)$. 
The values on $x$- and $y$-axis corresponds to integer position values 
$Mx$ and $Mp$ of Ulam cells in classical phase space 
($0\le x\le 1,\,0\le p\le 0.5$) 
for $M=400$ and $K=7$ (a), $K=5$ (b) and $K=2.5$ (c). 
The minimal resistance $R$ and the 
length of the optimal trajectory are $l=N_{E,A}(B)+\Delta l$
with $R=54.2721$, $N_{E,A}(B)=6$, $\Delta l=1$ (a); 
$R=46.8765$, $N_{E,A}(B)=8$, $\Delta l=0$ (b); 
$R=43.1982$, $N_{E,A}(B)=7$, $\Delta l=4$ (c).
}
\end{figure}

We consider that an intelligent surfer applies the above algorithm
to find the optimal short trajectory or path with minimal resistance
between nodes $A$ and $B$ of the Ulam network.
Thus, using the above efficient algorithm, based on the recomputing 
of the Erd\"os numbers of inverted dynamics (with respect to $B$ as a hub) 
and exploiting the condition (\ref{trajcond}), he/she computes 
the optimal trajectories with minimal resistance (\ref{resdef}) 
for networks with $200\le M\le 3200$ at $K=2.5,\,5,\,7$ 
and the initial/end points $A$/$B$ given in the previous Section..

In Fig.~\ref{fig1}, we show the obtained optimal trajectories for the three $K$ values 
and $M=400$ with a background color plot representing 
the set $S_{E,A}$ (of nodes $C$ with Erd\"os numbers 
$N_{E,A}(C)\le l_{\rm max}=N_{E,A}(B)+4$). We see, that all trajectory 
points are on these sets and in particular the initial values $i_j=1,2,3$ 
are in the limited orange regions of small Erd\"os numbers 
$N_{E,A}(C)=1,2,3$. Furthermore, for smaller $K=2.5$ ($K=5$) the set 
$S_{E,A}$ is strongly (modestly) reduced with respect to $K=7$. 
This is explained by the reduced values of $N_l\le 2\lceil 2+K\rceil$ 
requiring longer trajectories to cover the same fraction of phase space. 
The obtained values of the minimal resistance $R$  are of the same order of 
magnitude between the three $K$ values but for smaller $K$ there are 
somewhat smaller. This is due to the reduced $N_l$ values leading to an
increase of  typical transition probabilities and a resistance reduction 
(since $R$ is the sum of inverse transition probabilities).

\begin{figure}[t]
\begin{center}
  \includegraphics[width=0.46\textwidth]{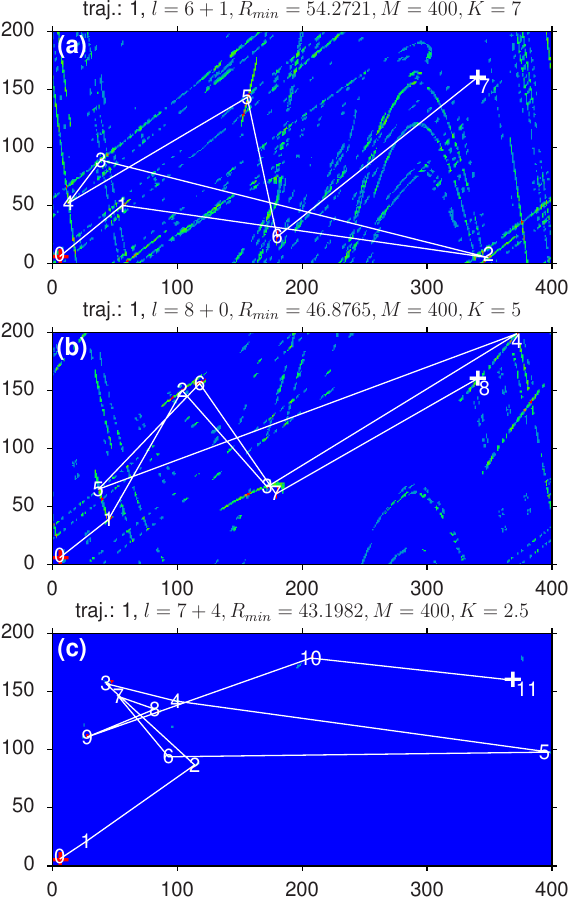}
\end{center}
\caption{\label{fig2}
As Fig.~\ref{fig1} with identical trajectories 
and same values of $K$ and $M$ 
but with a different background color plot showing the intersection 
set $S_I$ of cells $C$ with $N_{E,A}(C)+N^*_{E,B}(C)\le l_{\rm max}$
and where red/green/light blue/full blue corresponds 
to maximal/intermediate/small/negative (if $C\not\in S_I$) values of
the difference $l_{\rm max}-N_{E,A}(C)-N^*_{E,B}(C)$. 
}
\end{figure}

Fig.~\ref{fig2} shows the same trajectories but on a different background 
color plot representing the much smaller interaction set $S_I$ defined by the condition 
(\ref{trajcond2}). One sees that according to the condition (\ref{trajcond}) 
all trajectory points are indeed in set $S_I$. Actually, depending on the 
iteration number $j$ the trajectory points (for possible trajectories 
between $A$ and $B$) {\em cannot freely choose among all} points of $S_I$ since 
for each value of $j$ the condition (\ref{trajcond}) represents a subset 
of $S_I$ which is strictly smaller than $S_I$ if $j<l_{\rm max}$. 

We remind that the simple naive exponential search algorithm ensures 
automatically, by construction, that all search trajectories are in the 
(larger) set $S_{E,A}$ while the improved algorithm, exploiting the 
condition (\ref{trajcond2}), ensures that only certain search trajectories 
in the set $S_I$, that can indeed go the end point $B$, are used. 
The fact that the set $S_I$ is strongly reduced in comparison to $S_{E,A}$ 
illustrates the efficiency of the improved search algorithm. 
In fact, SupMat Fig.~S2 shows the number of points $N_I$ in the set $S_I$ versus 
network size $N\approx M^2/2$ and we see that $N_I<10^{4}$ and 
even $N_I\sim 10^{3}$ for the largest value of $M=3200$.
Thus a strong reduction of points which belong to the intersection
set  $S_I$ is at the origin of the efficiency of the intelligent surfer algorithm
based on the relations (\ref{trajcond}),(\ref{trajcond2}).

SupMat Fig.~S3 is similar to Figs.~\ref{fig1},\ref{fig2} but using 
as background color plot the set $S^*_{E,B}$ with limited inverse 
Erd\"os numbers (with respect to $B$ as hub). Now, the trajectory points 
close to the endpoint $B$ correspond to orange regions with small inverse 
Erd\"os number. Note that $S_I$ 
is a (strictly smaller) subset of the intersection $S_{E,A}\cap S^*_{E,B}$. 

\begin{figure}[H]
\begin{center}
  \includegraphics[width=0.46\textwidth]{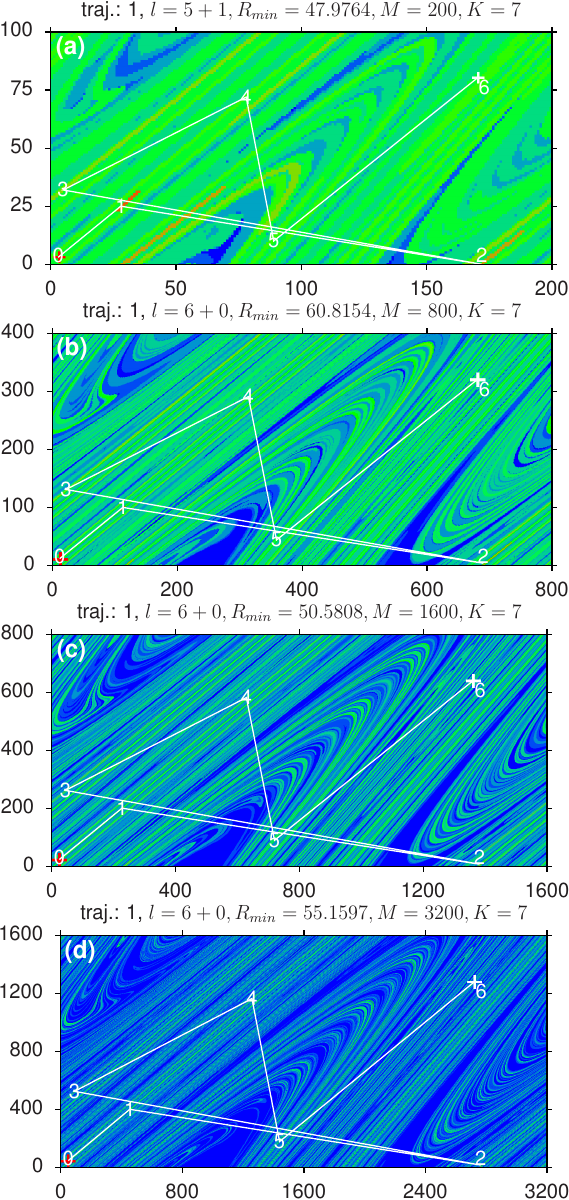}
\end{center}
\caption{\label{fig3}
As Fig.~\ref{fig1} with a background color plot showing the set 
$S_{E,A}$ for $K=7$ and $M=200$ (a), $M=800$ (b), $M=1600$ (c)
and $M=3200$ (d). The minimal resistance $R$ and the 
length of the optimal trajectory are respectively: $l=N_{E,A}(B)+\Delta l$
with $R=47.9764$, $N_{E,A}(B)=5$, $\Delta l=1$ (a);
$R=60.8154$, $N_{E,A}(B)=6$, $\Delta l=0$ (b); 
$R=50.5808$, $N_{E,A}(B)=6$, $\Delta l=0$ (c); 
$R=55.1597$, $N_{E,A}(B)=6$, $\Delta l=0$ (d).
}
\end{figure}

Fig.~\ref{fig3} is similar to Fig.~\ref{fig1} (with 
background color plot for the set $S_{E,A}$) but for the 
single value $K=7$ and four values $M=200,\,800,\,1600,\,3200$. 
The optimal trajectories for these cases are rather similar but a 
bit different from the optimal trajectory for $M=400$ visible 
in Fig.~\ref{fig1}(a). The fraction of nodes in the set $S_{E,A}$ 
decreases considerably with increasing $M$ and network size $N\approx M^2/2$ 
which is also due to the comparable values of $l_{\rm max}=9$ 
($M=200$) or $l_{\rm max}=10$ ($M\ge 400$). 

\begin{figure}[H]
\begin{center}
  \includegraphics[width=0.46\textwidth]{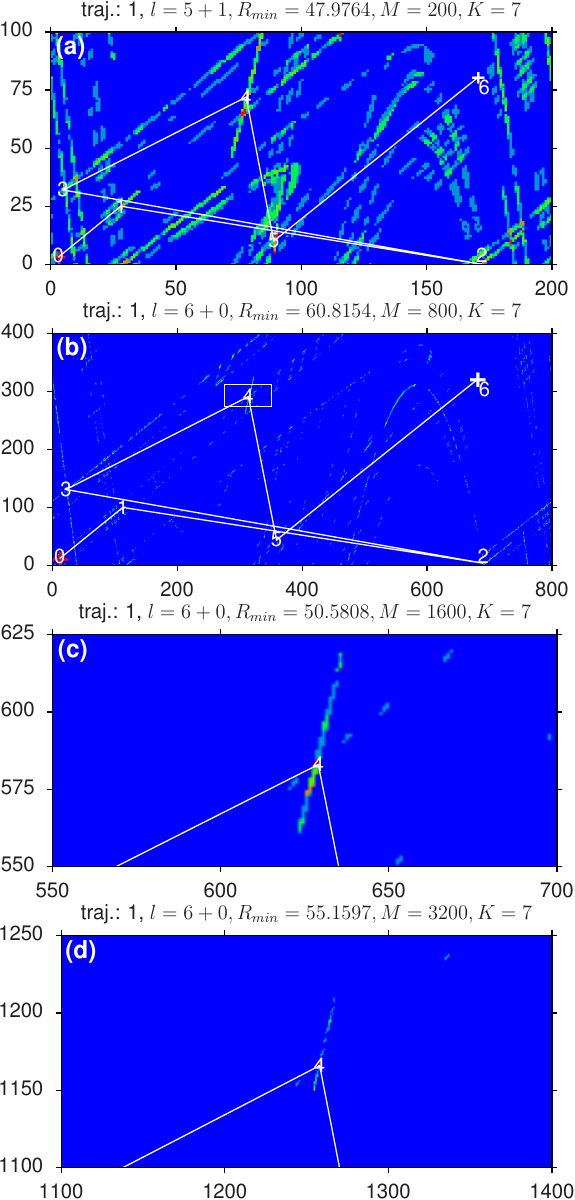}
\end{center}
\caption{\label{fig4}
As Fig.~\ref{fig3} with identical trajectories 
and same values of $K$ and $M$ 
but with a different background color plot showing 
the set $S_I$ (see caption of  Fig.~\ref{fig2}). 
The panels (c) ($M=1600$) and ($M=3200$) show a zoomed region 
of the phase space (around the point $i_4$) and 
corresponding to the white rectangle visible 
in panel (b) (case of $M=800$). 
}
\end{figure}

Fig.~\ref{fig4} is similar to Fig.~\ref{fig3} but with the intersection set $S_I$ as 
background color plot. Here the fraction of nodes in the set $S_I$ 
decreases very strongly with increasing $M$ and $N\approx M^2/2$.
In particular for $M\ge 1600$ only a zoomed region around the 4th node 
$i_4$ of the trajectory is shown (in full size plots the set $S_I$ 
would be essentially invisible for $M\ge 1600$ due to the limited 
resolution). Around $i_4$ the set of ``available points'' (for other 
trajectories from $A\to B$ with $l\le l_{\rm lmax}$) is rather well visible 
and of considerable size even though for larger values of $M$ also here 
a zoomed representation is necessary. 
The strong reduction of $S_I$ illustrates that the improved 
search algorithm gains more efficiency for larger values of $M$.

Results similar to Figs.~\ref{fig3},~\ref{fig4} are shown in 
SupMat Figs.~S4, S5 (for $K=5,\,2.5$ and with $S_{E,A}$ backgound) 
and Figs.~S6,~S7 (for $K=5,\,2.5$ and with $S_I$ backgound) 
For $K=5$ it turns out that the 
point $i_4$ is rather close to the end point $B$ such that the latter 
appears in the zoomed region for $M\ge 1600$. 
For $K=2.5$ the point $i_4$ of the optimal trajectory shifts considerably 
between $M=800$ and $M=1600$ such that $i_4$ at $M=800$ is outside 
the zoomed region used for $M\ge 1600$. 
Fig.~S8 of SupMat shows the 2nd, 8th and 10th best trajectory 
for $M=400$, $K=7$ using the background color plot of the set $S_I$. 
The three trajectories of Fig.~S8 and also the best trajectory visible 
in Fig.~\ref{fig2}(a) are somewhat different but the other 6 trajectories 
of the group of best 10 trajectories are rather close to one of those 4 
trajectories. 

\section{Fractal dimension of the set $S_I$}
\label{sec6}

\begin{figure}[H]
\begin{center}
  \includegraphics[width=0.46\textwidth]{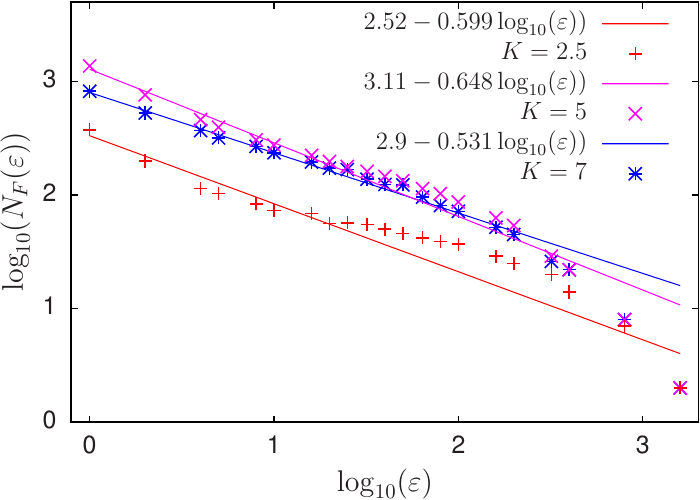}
\end{center}
\caption{\label{fig5}
Number $N_F(\eps)$ of boxes of size $\eps$ required to cover the set 
$S_I$ versus box size $\eps$ for $M=3200$, 
$K=7$ (blue $*$), $K=5$ (pink $\times$) and $K=2.5$ (red $+$). 
The values of $\eps$ are chosen such that $\eps=1,2,4,5,10,16,\ldots, M/2$ 
is a divisor of $M/2$. 
The straight lines correspond to the linear fit
$\log_{10}(N_F(\eps))=C-D_H\log_{10}(\eps)$ with 
$D_H=0.531\pm 0.013,\,C=2.9\pm 0.0087\,(K=7),\,$
$D_H=0.648\pm 0.02,\,C=3.11\pm 0.014\,(K=5)$ and
$D_H=0.599\pm 0.033,\,C=2.52\pm 0.022\,(K=2.5)$.
Here $D_H$ represents the box-counting fractal dimension and 
the fit has been done with a weight factor $w_j\sim 1/\eps_j$ for 
the different data points $(\eps_j,N_F(\eps_j))$ such that 
small $\eps$-values have a stronger weight. 
}
\end{figure}

In this Section we analyze the properties of
the set $S_I$. The small size of this set is
at the origin of the efficiency of the intelligent surfer algorithm
discussed above.
The color plots of the set $S_I$, shown in Figs.~\ref{fig2},~\ref{fig4}
for different cases, indicate a fractal structure of this set.
We therefore compute
the fractal box-counting dimension $D_H$ (see e.g. \cite{lichtenberg} and Refs. therein).
It is determined by the behavior 
$N_F(\eps)\sim\eps^{-D_H}$ where $N_F(\eps)$ is the number of boxes 
of size $\eps\times \eps$ necessary to cover the set $S_I$ (here
$\eps$ takes integer values when measured in units of the Ulam-grid).
Fig.~\ref{fig5} confirms the fractal behavior and shows, for $M=3200$, 
the linear fits $\log_{10}(N_F(\eps))=C-D_H\log_{10}(\eps)$ with 
$D_H=0.531\pm 0.013\,(K=7),\,$
$D_H=0.648\pm 0.02,\,(K=5)$ and
$D_H=0.599\pm 0.033,\,(K=2.5)$.
Thus the fractal dimension of the set $S_I$ is significantly smaller
the the phase-space dimension being 2 for this 2D symplectic map.
This facilitates the search of the optimal path between
the two points $A$ and $B$ for an intelligent surfer.

\section{Resistance dependence on network size}
\label{sec7}

\begin{figure}[H]
\begin{center}
  \includegraphics[width=0.46\textwidth]{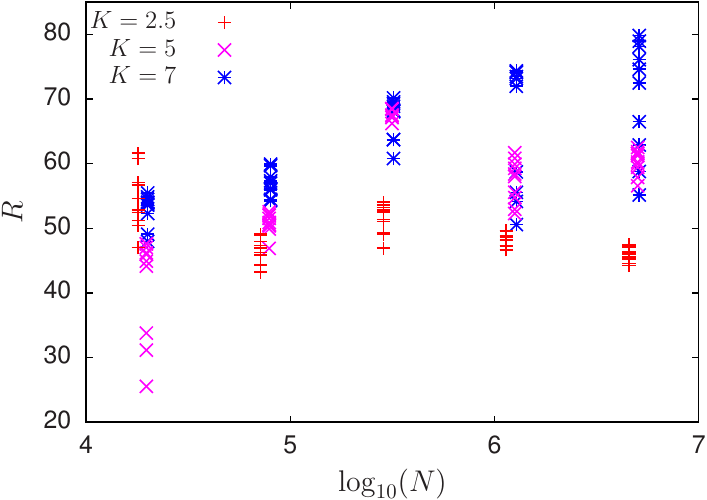}
\end{center}
\caption{\label{fig6}
Resistance $R$ of 10 optimal trajectories versus Ulam network size 
$N\approx M^2/2$ 
for $200\le M\le 3200$, $K=7$ (blue $*$), $K=5$ (pink $\times$) and $K=2.5$ 
(red $+$).
The resistance is defined as the sum of the inverse transition probabilities 
over each leg of the trajectory from $A$ to $B$ (see Eq. (\ref{resdef})). 
}
\end{figure}

It is important to determine the dependence of the 
resistance $R$ of the optimal path between $A$ and $B$
on the size $N$ of the Ulam network.
In Fig.~\ref{fig6} we show this resistance $R$ versus on $N \approx M^2/2 $
for the 10 best trajectories (with minimal resistance) and 
for all cases $2.5\le K\le 7$ and $200\le M\le 3200$.
The dependence of $R(N)$  
is not strictly monotone. For large $K$ values ($K=5, 7$)
there is a global tendency  that $R$ is growing
approximately logarithmically with $N$
($R \sim \log N$). This corresponds to a global
logarithmic growth of the typical  Erd\"os number with the 
Ulam network size $N$ discussed in \cite{smulam}.
For $K=2.5$ there are stability islands of significant size
and also the Lyapunov exponent is significantly
smaller in comparison to $K=5, 7$ \cite{chirikov}.
Due to this we expect that higher $N$ values
are required to see asymptotic dependence $R(N)$ in this case.
Also the presence of fluctuations between 10 optimal orbits
hides a slow logarithmic growth of $R(N)$.

We note that even for $N$ values bigger than a million
the values of $R$ remain relatively modest
with $R \sim 60$. This value
can be understood from the typical number of
links per node in the Ulam network.
Indeed due the map (\ref{eq_stmap}) structure
its maximal value
is approximately $N_l \approx 2(2+K) \approx 14 ; 18$
for $K=5; 7$ respectively \cite{smulam}.
We may estimate that a typical value
is smaller by a factor of 2
giving typically 10 links with a typical transition probability
$p_{i \rightarrow j} \sim 0.1$.
According to Figs.~\ref{fig1}-~\ref{fig4}
it takes about $l_{\rm opt} \approx 6$ transitions
for an optimal (or quasi-optimal) path
to go from $A$ to $B$.
Thus a typical resistance is $R \sim l_{\rm opt}/p_{i \rightarrow j} \sim 60$
in agreement with the results of Fig.~\ref{fig6}.

\section{Case of naive surfer}
\label{sec8}

In order to compare the results of the previous Sections 
for an {\em intelligent surfer}, who can apply the optimal algorithm to 
find (i) efficiently trajectories from $A\to B$ and (ii) those with 
minimal resistance, we consider in this Section a different model 
of a {\em naive surfer}. This naive surfer starts at the initial point $i_0=A$ 
and then searches among all nodes accessible from $A$ the cell $i_1$ 
which has the minimal Euclidean distance 
($\sqrt{(p_{i_1}-p_B)^2+(x_{i_1}-x_B)^2})$ 
in phase space to the target node $B$. Then from this node $i_1$ he 
searches the accessible node $i_2$ with minimal Euclidean distance to $B$ etc.
If at some point some $i_j$ he can reach the final node $B$ with one step 
the algorithm will automatically terminate with a found trajectory. 
A priori, one expects that the time scale for finding the end point $B$ is 
$\sim N$. 

However, due to the birthday paradox (see e.g. \cite{birth} and Refs. therein)
the typical time scale when this naive 
surfer comes back to an already visited node is $\sim \sqrt{N}\ll N$ and 
when this happens he enters in a periodic trajectory which will never 
reach the end point $B$. Therefore, the naive surfer should also keep a list 
of all visited nodes and as next step he only chooses nodes with minimal 
distance to $B$ among non-visited nodes yet. Even in this case there may be 
an ``accident'' when at a certain position all accessible nodes have 
already been visited. 
In this case, he goes back one step and choose another position 
(with minimal distance and not yet visited). 

We have implemented the algorithm of this naive surfer and it turns 
that the first revisited nodes indeed happens at an iteration time scale 
$j_{\rm 1st\ revisited}\sim \sqrt{N}$. Also the 2nd type of accident 
happens a few times but in these cases only one simple back step is 
necessary to get out of the periodic loop. The results are given in SupMat Figs.~S9,~S10.

SupMat Fig. S9 shows two such trajectories for $K=5$ and $M=200,400$ 
with colors from blue to red representing small to large iteration numbers. 
In both cases a finite fraction of all available cells/nodes is used. 
The structure of the sets of trajectory points is quite random with 
no visible phase space structure (apart from the hole due to a quite big 
stable island at $(x,p)\approx (0.6,\,0.3)$).
SupMat Fig. S10 shows the length $l_{\rm naive}$ of these trajectories 
versus network size $N$ (for all 15 values of $(K,M)$).

In all studied cases 
$l_{\rm naive}$ is in the interval $[0.1\times N,\,N]$
thus being of the order of network size $N$.
Obviously this kind of strategy to find short trajectories to the end point $B$, 
without using inverse Erd\"os numbers, is very inefficient.
Furthermore, the above algorithm of a naive surfer cannot find the trajectory with 
minimal resistance.

\section{Intelligent surfer with limited resources}
\label{sec9}

Ideally an intelligent surfer could reproduce the algorithm presented above 
(in Section IV) exploiting at each search level the condition 
(\ref{trajcond}) to 
avoid as early as possible all trajectories that do not link the 
two cells $A$ and $B$. For this he needs to compute the values 
of inverse Erd\"os numbers with respect to $B$ as hub at least up to the level 
$l_{\rm max}$, i.e. to determine $N^*_{E,B}(C)$ for all cells $C$ 
with $N^*_{E,B}(C)\le l_{\rm max}$. 
In principle, he also needs the regular Erd\"os number $N_{E,B}(C)$ 
(with respect to the hub $A$ and the non-inverted initial Ulam network) 
at least of the node $C=B$ to determine the value of 
$l_{\rm max}=N_{E,A}(B)+4$.

Let us now assume that the surfer disposes only of limited resources 
(for example in storage of Erd\"os numbers) 
and tries to minimize his initial efforts to compute both types of 
Erd\"os numbers. Concerning the computation of the value $N_{E,A}(B)$
(the only value of regular Erd\"os numbers which is needed) 
he could try to choose ad hoc some small value of 
$l_{\rm max}$ and apply the above algorithm which is highly efficient 
especially for small values of $l_{\rm max}$. If $l_{\rm max}\ge N_{E,A}(B)$ 
he will find the non-empty set of all trajectories between $A$ and $B$ 
and therefore also the minimal length $l$ of these trajectories which 
is just $N_{E,A}(B)$. If $l_{\rm max}< N_{E,A}(B)$ he will find no solutions 
and in this case he can increase $l_{\rm max}$ by one or some small value 
and try again.

However, in order to use the above algorithm he still needs to recompute 
the inverse Erd\"os numbers for a large set of nodes $C$ up to level 
$l_{\rm max}$. The simpler and more expensive algorithm (without the 
test of the condition (\ref{trajcond})) becomes most expensive in 
the final steps where the better algorithm only uses small inverse 
Erd\"os numbers. 
Therefore the intelligent surfer, could decide to compute the 
set of inverse Erd\"os numbers only up to a limited depth $l_S<l_{\rm max}$ 
which is much less expensive since $N_l^{l_S}\ll N_l^{l_{\rm max}}$ 
and to apply the test of the condition (\ref{trajcond}) only in the 
final steps when $j\ge l_{\rm max}-l_S$ which requires only 
to know $N^*_{E,B}(C)$ up to level $l_S$. 

In order to measure the price/efficiency of such a simplified algorithm, 
we perform the recursive search in this way and we compute the sum 
of three quantities which are (i) the number $N_T$ of found trajectories between 
$A$ and $B$, (ii) the number of times 
the recursion is stopped because the condition (\ref{trajcond}) is 
not verified (if $j\ge l_{\rm max}-l_S$) and (iii) the number of times 
the recursion goes to the level $l_{\rm max}$ but without finding a trajectory 
containing the final point $B$ 
(the case (iii) actually only happens if $l_S=0$). 
These three cases correspond to the three possibilities to stop the search 
recursion and their total number, which we call the {\em search effort number} 
$N_S$, corresponds very accurately to the computational effort. We have 
verified for a few cases that $N_S$ is indeed proportional to the exact 
computation time for different values of $l_S$. 
In the case of the perfect algorithm (with $l_S=l_{\rm max}$) it turns 
out that $N_S$ is typically $(10-22)\times N_T$ which is larger than $N_T$ 
but only by one order of magnitude showing that the perfect algorithm is 
very efficient to find rather directly the $N_T$ trajectories 
between $A$ and $B$. 

\begin{figure}[H]
\begin{center}
  \includegraphics[width=0.46\textwidth]{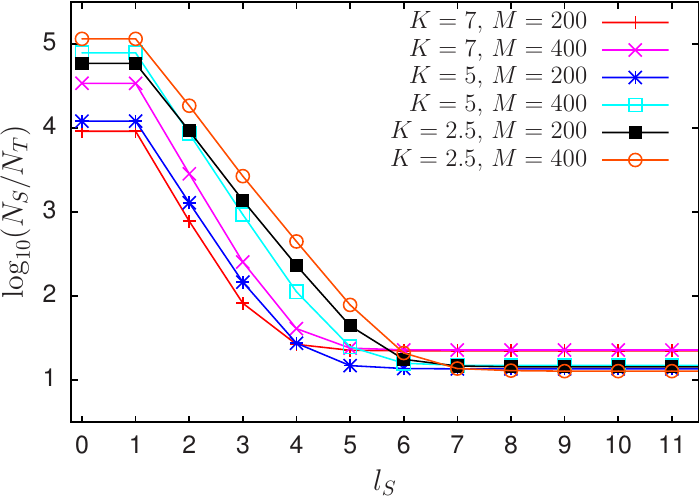}
\end{center}
\caption{\label{fig7}
Shown is the (logarithm of the) ratio of the search effort 
number $N_S$ with the number $N_T$ of possible trajectories between $A$ and 
$B$ versus search parameter $l_S$ (see text) for $K=2.5,\,5,\,7$ and 
$M=200,\,400$.
}
\end{figure}

We have computed $N_S$ for different values of $l_S$, $K=2.5,\,5,\,7$ 
and small values 
of $M=200,\,400$ (for larger values of $M$ the cases $l_S=0$ and $l_S=1$ 
become very expensive). The result is shown in Fig.~\ref{fig7} showing 
(the logarithm) of the ratio $N_S/N_T$ versus $l_S$ which starts 
at typical values $N_S/N_T=10^4$-$10^5$ and decays very rapidly 
to a saturation value $N_S/N_T=10$-$22$ for $l_S\ge 6$-$7$.
Here the value $N_T$ of possible trajectories for the above parameters 
depend on $K$ and $M$ and is in the range $10^{3}<N_T<10^7$ (see SupMat Fig.~S1 
and here the two sets of data points with $N<10^5$ corresponding 
to $M=200$ or $M=400$ respectively). 
Actually, at $l_S=3$-$4$, we already have a reduction of $N_S$ by 
a factor of 100 reducing the effort to 1\% showing that the strategy 
of the intelligent surfer to limit the value of $l_S$ is indeed liable.

SupMat Figure~S11 shows the (logarithm of the) difference 
$N_S-N_{S,{\rm min}}$ where $N_{S,{\rm min}}$ is the minimal value 
at $l_S=l_{\rm max}$. This representation amplifies the small differences 
when $N_S$ has nearly converged to $N_{S,{\rm min}}$. At certain 
critical values $l_S=8$-$12$ we have $N_S=N_{S,{\rm min}}$ and 
Fig.~S11 does not show data points for these cases (due to the 
logarithm). In contrast to Fig.~\ref{fig7} the convergence to the limiting 
value seems a bit later but this is artificial due to the different 
presentation. The main interpretation of Fig.~S11 is finally 
the same as in Fig.~\ref{fig7}: at $l_S=3$-$4$ there is already 
a strong reduction of $N_S$ and the computational effort. 

One could ask the question which is the optimal value of $l_S$ to minimize 
the global computational cost of both pre-computation of 
partial inverse Erd\"os numbers of level below $l_S$ and 
the initial exponential search algorithm. The limited 
pre-computation of inverse Erd\"os numbers costs roughly 
$\sim N_l^{l_S}$ operations and the search algorithm needs 
mostly $\sim N_l^{l_{\rm max}-l_S}$ since it is essentially exponential 
for $l<l_{\rm max}-l_S$. The total cost 
\begin{equation}
\label{totalcost}
N_C(l_S)=N_l^{l_S}+N_l^{l_{\rm max}-l_S}
\end{equation}
is obviously minimal at $l_S=l_{\rm max}/2\approx 5$-$6$ for our 
parameters used above and close the value $l_S\approx 4$ which gives
a considerable reduction of $N_S$ according to Fig.~\ref{fig7}.

\section{Discussion}
\label{sec10}

We have considered a surfer moving in a chaotic flow
who is facing a goal quest to determine optimal 
Ulam network trajectories with minimal resistance between an initial point $A$
and another final point $B$. The Ulam network is 
generated from the symplectic Chirikov standard map with dynamical chaos 
and the Perron-Frobenius eigenvector with maximal eigenvalue
has ergodic probability equipartition over cells (nodes) that
belong to one big connected chaotic component. 

We propose an algorithm 
for an intelligent surfer which allows to
find a requested path of minimal resistance
with a complexity $N_l^{l_{\rm max}/2}$ where $N_l$ is the typical 
number of links per network node and $l_{\rm max}$ the maximal length 
of considered trajectories. 
Due to the exponential Lyapunov instability
of chaotic dynamics the number of required transitions $l\le l_{\rm max}$
grows only logarithmically as $l \sim \log N$ with the network size $N$. 

The efficiency of the algorithm is based
on the computation of Erd\"os numbers for the 
directed and time inverted dynamical flow.
In particular, we have constructed the intersection set $S_I$ 
of nodes satisfying the condition (\ref{trajcond2}) and 
having a fractal dimension $0.5$-$0.6$ being 
significantly smaller than the phase space dimension $2$. 

The developed algorithm for the intelligent surfer is 
exponentially more efficient compared to a case of a naive surfer, 
who tries at each step to minimized the distance to the target point $B$, 
or to a direct recursive search algorithm without using the interaction 
set $S_I$. 

 We mention that most of published works and algorithms (see e.g. 
\cite{path1,path2,path3,path4}) concern typically 
directed networks and shortest trajectories where the iteration length $l$ is 
minimized and not the resistance $R$ (\ref{resdef}). In our language the 
computation of the minimal iteration length $l$ is equivalent to the computation 
of the Erd\"os number. However, when a minimization of resistance is required,
a more complex algorithm, based on the precomputation
of inverse Erd\"os numbers, is needed to be used. 

We hope that the algorithm of an intelligent surfer we propose will allow to perform
a control of motion in the regime of developed chaos in a better and more efficient way.

$\;\;$\\

 {\bf Acknowledgments:}
\bigskip

This work has been partially supported through the grant
NANOX $N^o$ ANR-17-EURE-0009 in the framework of 
the Programme Investissements d'Avenir (project MTDINA).
This work was granted access to the HPC resources of 
CALMIP (Toulouse) under the allocation 2023-P0110.


\clearpage



\setcounter{figure}{0} \renewcommand{\thefigure}{S\arabic{figure}} 
\setcounter{equation}{0} \renewcommand{\theequation}{S.\arabic{equation}} 
\setcounter{page}{1}

\noindent{{\bf Supplementary Material for\\
\vskip 0.2cm
Goal quest for an intelligent surfer\\
moving in a chaotic flow}
\bigskip

\noindent by
K.~M.~Frahm and D.~L.~Shepelyansky\\
\noindent Laboratoire de Physique Th\'eorique, 
Universit\'e de Toulouse, CNRS, UPS, 31062 Toulouse, France
\bigskip

Here we present more data and Figures for the main part of the article.

\begin{figure}[H]
\begin{center}
  \includegraphics[width=0.46\textwidth]{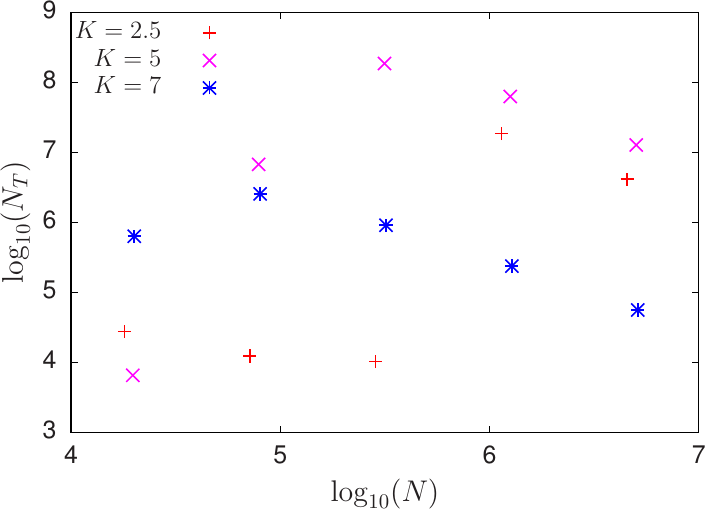}
\end{center}
\caption{\label{figS1}
Number $N_T$ of possible trajectories from $A$ to $B$ 
with length $l\le l_{\rm max}= N_{E,A}(B)+4$
versus Ulam network size $N\approx M^2/2$ 
for $200\le M\le 3200$, $K=7$ (blue $*$), $K=5$ (pink $\times$) and $K=2.5$ 
(red $+$).
}
\end{figure}

\begin{figure}[H]
\begin{center}
  \includegraphics[width=0.46\textwidth]{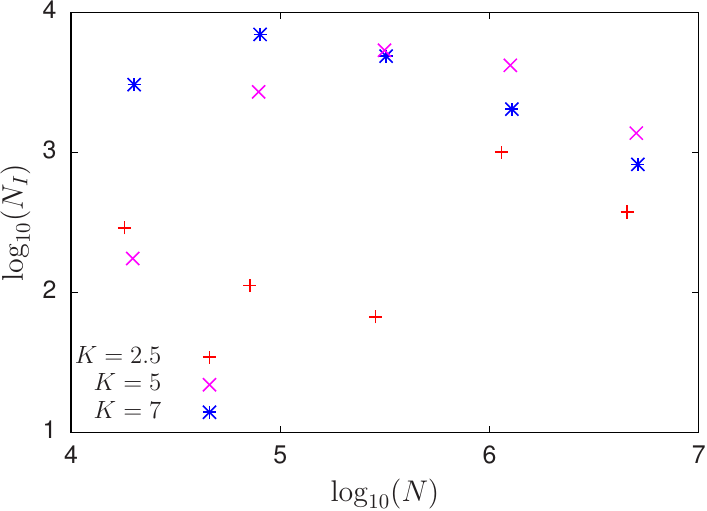}
\end{center}
\caption{\label{figS2}
Size $N_I$ of the set $S_I$ versus the size of Ulam network $N\approx M^2/2$ 
for $200\le M\le 3200$, $K=7$ (blue $*$), $K=5$ (pink $\times$) and $K=2.5$ 
(red $+$). Note that $N_I$ corresponds to $N_F(\eps=1)$ 
and the three data points at $M=3200$ ($\lg_{10}(N) \approx 6.7$) correspond to 
the three data points at $\log_{10}(\eps)=0$ in Fig. 5.
}
\end{figure}

\begin{figure}[H]
\begin{center}
  \includegraphics[width=0.46\textwidth]{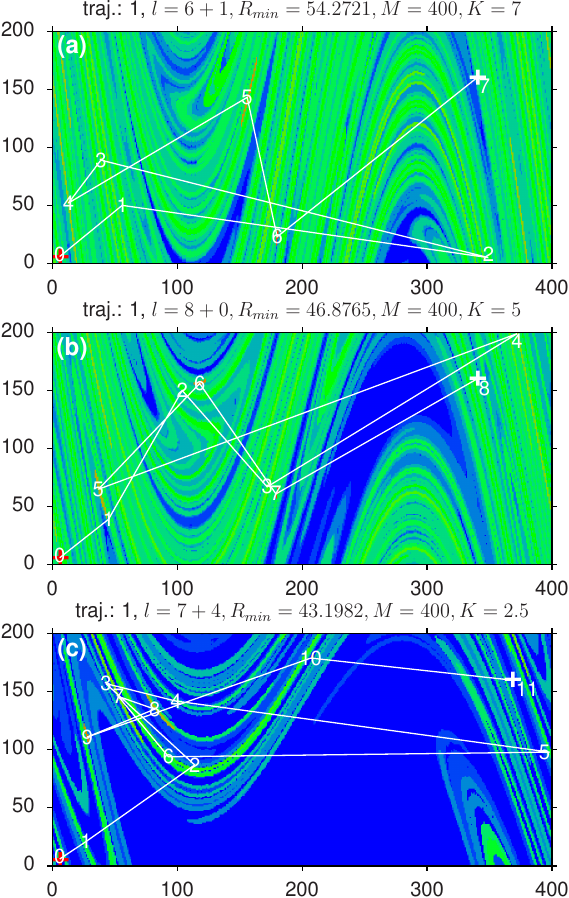}
\end{center}
\caption{\label{figS3}
As Figs.~1, 2 with identical trajectories 
and same values of $K$ and $M$ 
but with a different background color plot showing 
the set $S^*_{E,A}$ of cells $C$ with $N^*_{E,B}(C)\le l_{\rm max}$
and where red/green/light blue/full blue corresponds 
to maximal/intermediate/small/negative (if $C\not\in S^*_{E,A}$) values of
the difference $l_{\rm max}-N^*_{E,B}(C)$. 
}
\end{figure}
\begin{figure}[H]
\begin{center}
  \includegraphics[width=0.46\textwidth]{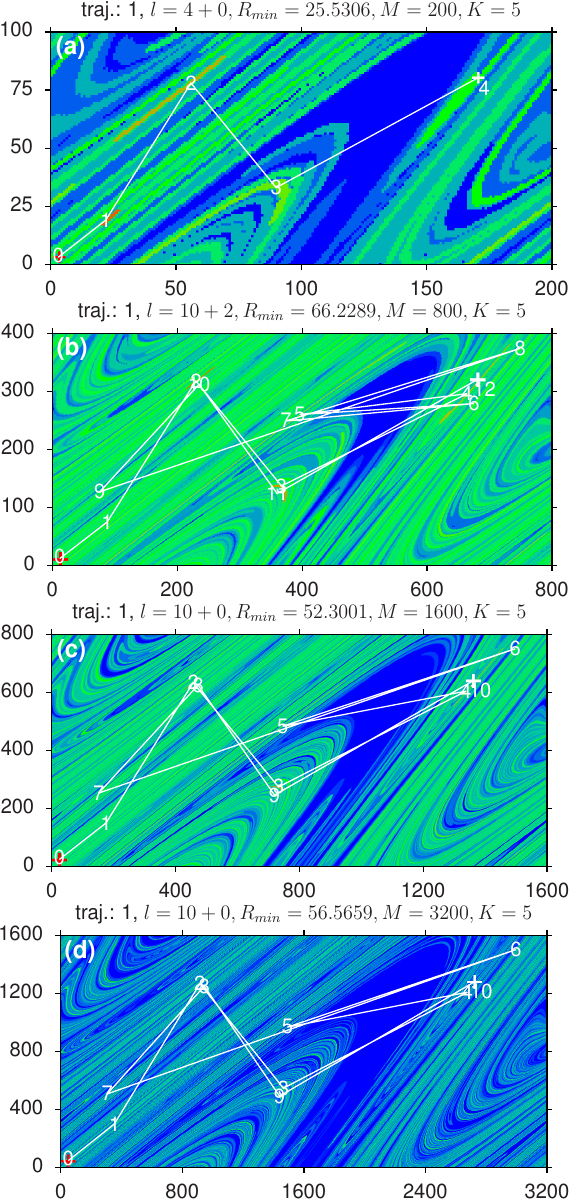}
\end{center}
\caption{\label{figS4}
As Fig.~3 with background showing the set $S_{E,A}$ 
but for $K=5$, same $M$ values. The minimal resistance $R$ and the 
length of the optimal trajectory are $l=N_{E,A}(B)+\Delta l$
with $R=25.5306$, $N_{E,A}(B)=4$, $\Delta l=0$ (a);
$R=66.2289$, $N_{E,A}(B)=10$, $\Delta l=2$ (b); 
$R=52.3001$, $N_{E,A}(B)=10$, $\Delta l=0$ (c); 
$R=56.5659$, $N_{E,A}(B)=10$, $\Delta l=0$ (d).
}
\end{figure}
\begin{figure}[H]
\begin{center}
  \includegraphics[width=0.46\textwidth]{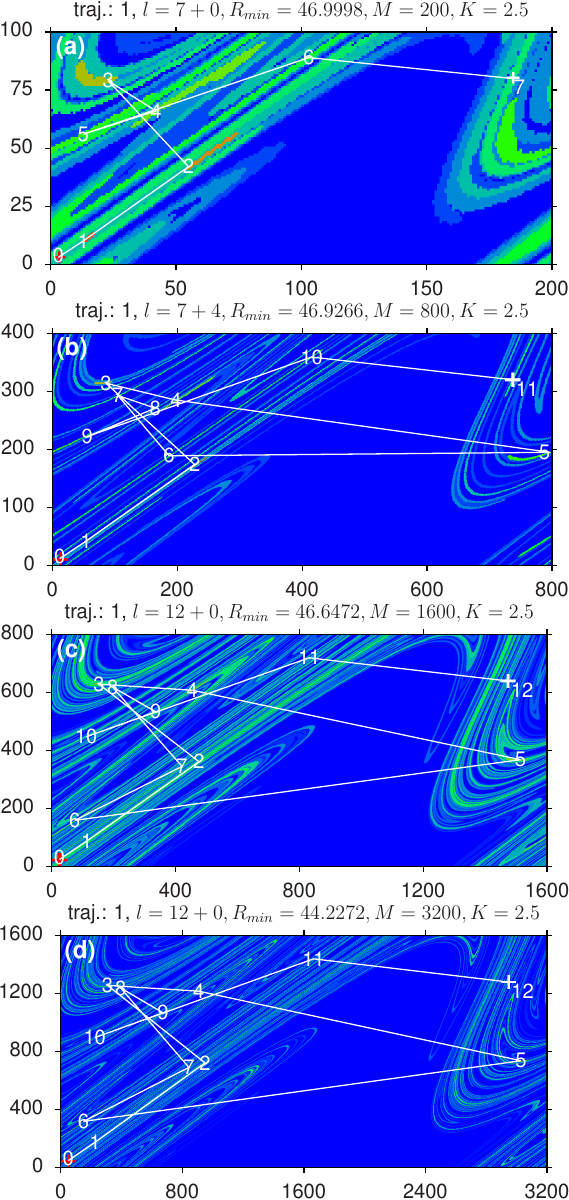}
\end{center}
\caption{\label{figS5}
As Fig.~3 with background showing the set $S_{E,A}$ 
but for $K=2.5$, same $M$ values. The minimal resistance $R$ and the 
length of the optimal trajectory are $l=N_{E,A}(B)+\Delta l$
with $R=46.9998$, $N_{E,A}(B)=7$, $\Delta l=0$ (a);
$R=46.9266$, $N_{E,A}(B)=7$, $\Delta l=4$ (b); 
$R=46.6472$, $N_{E,A}(B)=7$, $\Delta l=4$ (c); 
$R=44.2272$, $N_{E,A}(B)=12$, $\Delta l=0$ (d).
}
\end{figure}
\begin{figure}[H]
\begin{center}
  \includegraphics[width=0.46\textwidth]{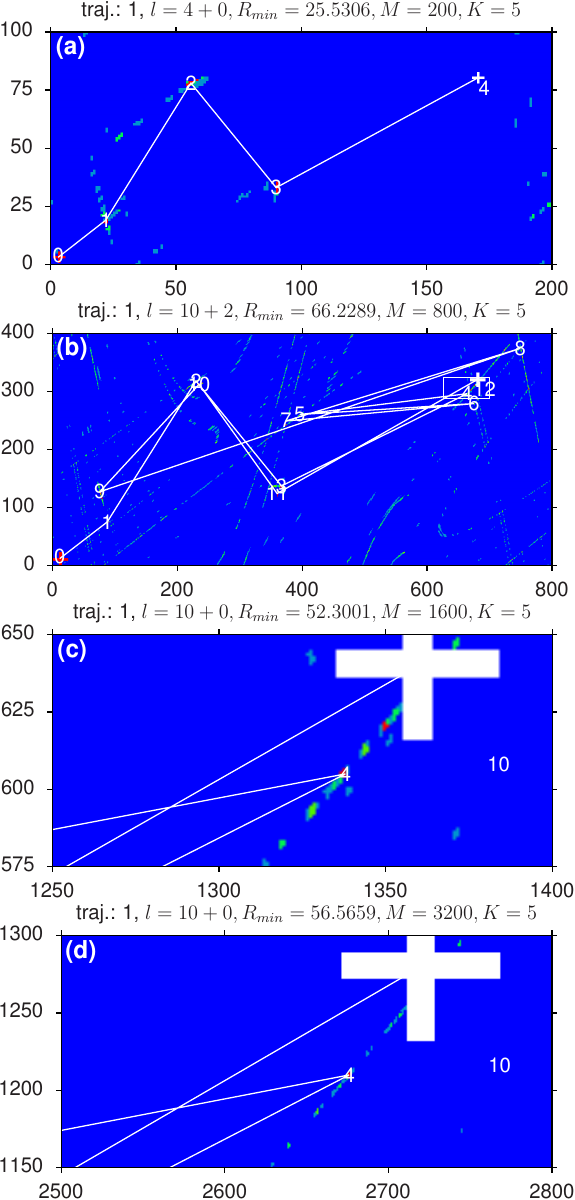}
\end{center}
\caption{\label{figS6}
As Fig.~4 with background showing the set $S_I$ 
but for $K=5$, same $M$ values, same trajectories of Fig.~4. 
As in Fig.~4 the panels (c) ($M=1600$) and ($M=3200$) 
show a zoomed region (around the point $i_4$) 
of the phase space corresponding to the white 
rectangle visible in panel (b) (case of $M=800$). 
The white number $10$ refers to the iteration number of the visible 
end point $B=i_{10}$ (big white cross) which also happens to be inside 
the zoomed rectangle. 
}
\end{figure}
\begin{figure}[H]
\begin{center}
  \includegraphics[width=0.46\textwidth]{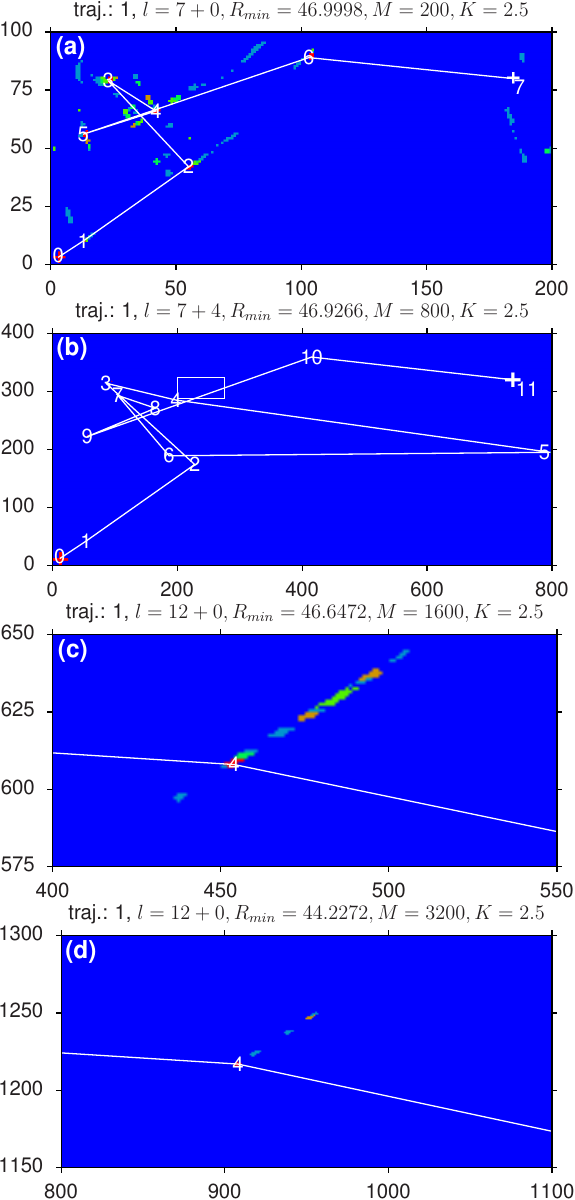}
\end{center}
\caption{\label{figS7}
As Fig.~4 with background showing the set $S_I$ 
but for $K=2.5$, same $M$ values, same trajectories of Fig.~\ref{figS5}.
As in Fig.~4 the panels (c) ($M=1600$) and ($M=3200$) 
show a zoomed region (around the point $i_4$ at given $M$-value) 
of the phase space corresponding to the white 
rectangle visible in panel (b) (case of $M=800$). 
Note that here the point $i_4$ of the optimal trajectory at $M=800$ (b)
has a different position (outside the white rectangle) 
as compared to the two cases $M=1600$ (c) and $M=3200$ (d).
}
\end{figure}

\begin{figure}[H]
\begin{center}
  \includegraphics[width=0.46\textwidth]{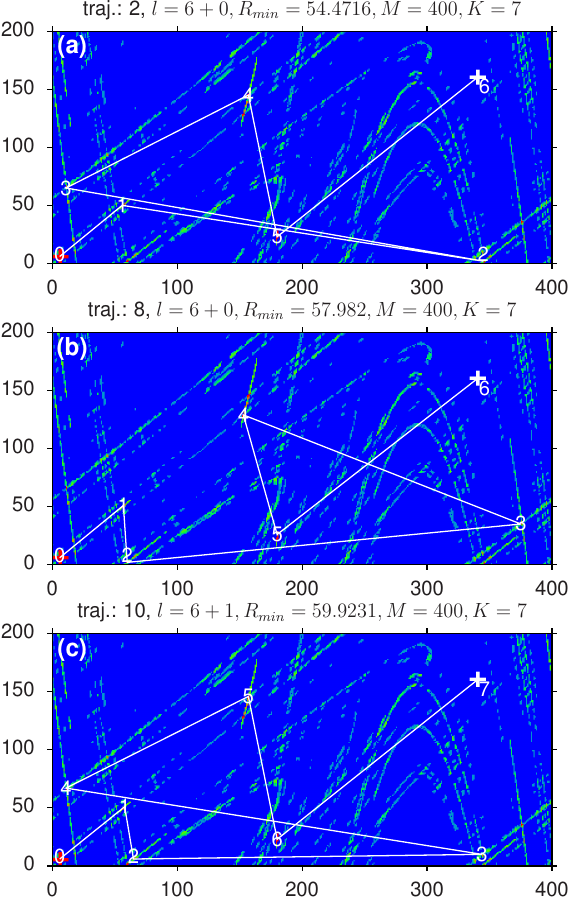}
\end{center}
\caption{\label{figS8}
As Fig.~2 with background showing the set $S_I$ for 
$K=7$, $M=400$, but for the 2nd (a), 8th (b) and 10th (c) best 
trajectory with respect to minimal $R$. 
The minimal resistance $R$ and the 
length of the trajectories are $l=N_{E,A}(B)+\Delta l$
with $R=54.4716$, $N_{E,A}(B)=6$, $\Delta l=1$ (a);
$R=57.982$, $N_{E,A}(B)=6$, $\Delta l=0$ (b); 
$R=59.9231$, $N_{E,A}(B)=6$, $\Delta l=0$ (c).
}
\end{figure}

\begin{figure}[H]
\begin{center}
  \includegraphics[width=0.46\textwidth]{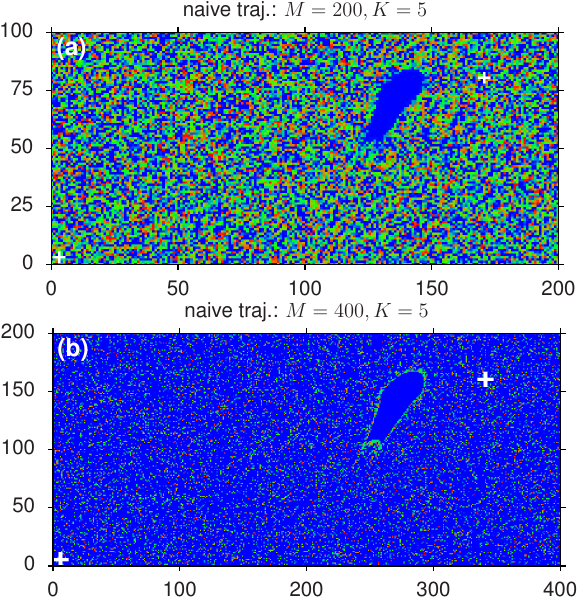}
\end{center}
\caption{\label{figS9}
Trajectory of the naive surfer for $K=5$, $M=200$ (a) and 
$M=400$ (b). The colors red (green, light blue) indicate 
cells with iteration numbers which are maximal (intermediate, 
small) in comparison to the length $l_{\rm naive}$ of the trajectory. 
Full blue indicate cells which are not visited by the trajectory and also
stable islands whose cells are not present in the Ulam network.
The white crosses show the positions of the initial cell A (bottom left 
corner) and the final cell (top right corner). 
The values of $l_{\rm naive}$ are 13982 (a) or 13744 (b) 
representing the fraction $l_{\rm naive}/N=70.96\%$ (a), 
$17.46\%$ (b). 
}
\end{figure}

\begin{figure}[H]
\begin{center}
  \includegraphics[width=0.46\textwidth]{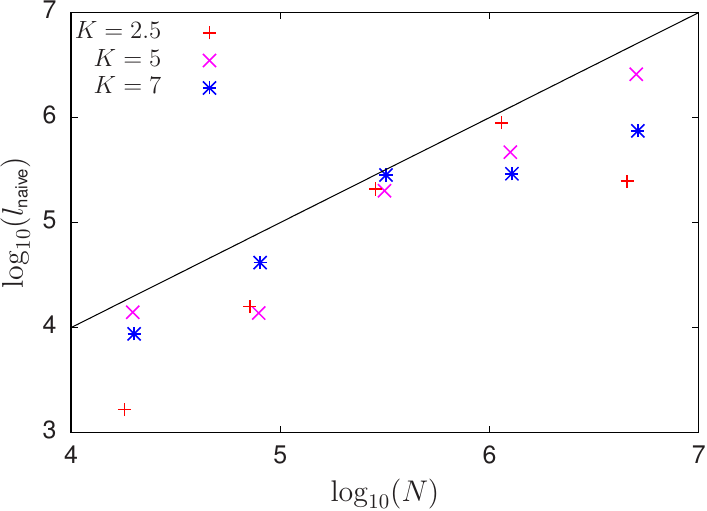}
\end{center}
\caption{\label{figS10}
Length $l_{\rm naive}$ of the trajectory of the naive surfer 
versus Ulam network size $N\approx M^2/2$ 
for $200\le M\le 3200$, $K=7$ (blue $*$), $K=5$ (pink $\times$) and $K=2.5$ 
(red $+$). The straight black line shows for comparison the case 
$l_{\rm naive}=N$.
}
\end{figure}

\begin{figure}[H]
\begin{center}
  \includegraphics[width=0.46\textwidth]{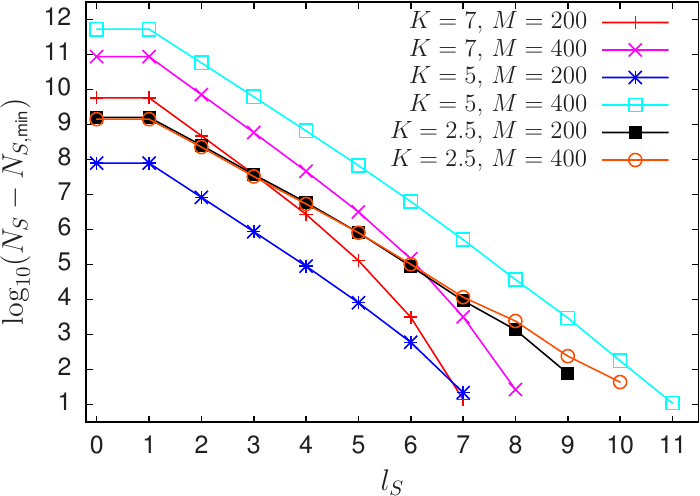}
\end{center}
\caption{\label{figS11}
Shown is the (logarithm of the) difference of the search effort 
number $N_S$ with its minimal value $N_{S,{\rm min}}$ versus search 
parameter $l_S$ (see text) for $K=2.5,\,5,\,7$ and $M=200,\,400$.
Note that in contrast of Fig.~7 the data points 
where $N_S=N_{S,{\rm min}}$, for sufficiently 
large $l_S$, are not visible since the logarithm of zero is not defined. 
}
\end{figure}


\begin{thebibliography}{99}

\bibitem{lichtenberg} A. J. Lichtenberg, M. A. Lieberman, 
         {\it Regular and chaotic dynamics}, Springer, Berlin (1992).
\bibitem{cvitanovic} P.~Cvitanovic, R.~Artuso, R.~Mainieri, G. Tanner, G.~Vattay,
        N.~Whelan and A.~Wirsba,
        {\it Chaos: Classical and Quantum},
        http://chaosbook.org/ (accessed May 2023).
 \bibitem{ulam} S.M.~Ulam, {\it A Collection of mathematical problems},
        Vol. 8 of {\it Interscience tracs in pure and 
        applied mathematics},
        Interscience, New York, p. 73 (1960).
\bibitem{li} T.-Y.~Li,  
        {\it Finite approximation for the Perron-~Frobenius operator, 
        a solution to Ulam's conjecture},
        J. Approx. Theory {\bf 17}, 177 (1976).
\bibitem{tel} Z.~Kov\'acs and T.~T\'el,
          {\it Scaling in multifractals: discretization 
         of an eigenvalue problem},
         Phys. Rev. A {\bf 40}, 4641 (1989).
\bibitem{kaufmann} Z.~Kaufmann, H.~Lustfeld, and J.~Bene,
         {\it Eigenvalue spectrum of the Frobenius~-Perron operator near intermittency},
         Phys. Rev. E {\bf 53}, 1416 (1996).
\bibitem{froyland2007} G.~Froyland, R.~Murray, and D.~Terhesiu,
        {\it   Efficient computation of topological entropy, pressure,
        conformal measures, and equilibrium states in one dimension},
        Phys. Rev. E {\bf 76}, 036702 (2007).
\bibitem{ding} J.~Ding and A.~Zhou,
         {\it Finite approximations of Frobenius~-Perron operators:
         A solution of Ulam’s conjecture to multidimensional transformations},
         Physica D {\bf 92}, 61 (1996).
\bibitem{liverani} M.~Blank, G.~Keller, and C.~Liverani,
        {\it Ruelle~-Perron~-Frobenius spectrum for Anosov maps},
        Nonlinearity {\bf 15}, 1905 (2002).
\bibitem{froyland2008a} D.~Terhesiu and G.~Froyland,
        {\it Rigorous numerical approximation of
         Ruelle--Perron--Frobenius
        operators and topological pressure of expanding maps},
        Nonlinearity {\bf 21}, 1953 (2008).
\bibitem{froyland2008b} G.~Froyland, S.~Lloyd, and A.~Quas,
         {\it Coherent structures and isolated spectrum for Perron–Frobenius cocycles},
         Ergod. Th. Dynam. Sys. {\bf 30(3)}, 729 (2010).
\bibitem{chirikov} B.~V.~Chirikov, {\it A universal instability 
        of many-dimensional  oscillator systems}, 
        {\em  Phys. Rep.} {\bf 52} (1979) 263.
\bibitem{frahmulam} K. M. Frahm and D. L. Shepelyansky,
        {\it Ulam method for the Chirikov standard map},
        Eur. Phys. J. B {\bf 76}, 57 (2010).
\bibitem{ulampoincare} K. M.~Frahm and  D. L.~Shepelyansky,
         {\it Poincar\'e recurrences and Ulam method for the Chirikov standard map},
           Eur. Phys. J. B {\bf 86}, 322 (2013).
\bibitem{smulam} K. M.~Frahm and  D. L.~Shepelyansky,
         {\it Small world of Ulam networks for chaotic Hamiltonian dynamics}, 
           Phys. Rev. E {\bf 98}, 032205 (2018).
\bibitem{milgram} S.~Milgram,
         {\it The small-world problem},
         Psychology Today {\bf 1(1)}, 61  (May 1967).
\bibitem{strogatz} D. J.~Watts and S. H.~Strogatz,
          {\it Collective dynamics
           of 'small-world' networks},
          Nature {\bf 393}, 440 (1998).
\bibitem{newman} M. E. J.~Newman, 
           {\it The structure and function of complex networks},
           SIAM Rev. {\bf 45(2)}, 167 (2003).
\bibitem{dorogovtsev} S.~Dorogovtsev,
         {\it Lectures on complex networks},
         Oxford University Press, Oxford (2010). 
\bibitem{erdos} P.~Erd\"os and A.~R\'enyi, 
         {\it On random graphs I},
         Publicationes Mathematicae {\bf 6}, 290 (1959).
\bibitem{brin} S.~Brin and L.~Page, 
         {\it The anatomy of a large-scale hypertextual Web search engine},
         Computer Networks and 
         ISDN Systems {\bf 30}, 107 (1998)
\bibitem{meyerbook} A.~M.~Langville and C.~D.~Meyer, {\it Google's
         PageRank and Beyond: The Science of  Search Engine Rankings},
         Princeton University Press, Princeton (2006).
\bibitem{rmp2015} L.~Ermann, K. M. ~Frahm and D. L.~Shepelyansky,
          {\it Google matrix analysis of directed networks},
          Rev. Mod. Phys. {\bf 87}, 1261 (2015).
\bibitem{path1} E. W.~Dijkstra,
         {\it A note on two problems in connexion with graphs},
         Numerische Mathematik {\bf 1}, 269 (1959).
\bibitem{path2}  S. E.~Dreyfus,
          {\it An appraisal of some shortest-path algorithms},
          Operational Research {\bf  17(3)}, 373 (1969).
\bibitem{path3} D. B.~Johnson,
         {\it Efficient algorithms for shortest Ppaths in sparse networks},
         J. Association Comp. Machinery {\bf24(1)}, 1 (1977).
\bibitem{path4} M. E. J.~Newman,
         {\it  Scientific collaboration networks. II.
         Shortest paths, weighted networks, and centrality},
         Phys. Rev. E {\bf 64}, 016132 (2001).
\bibitem{stmapscholar} B.~Chirikov and D.~Shepelyansky, 
         {\it Chirikov standard map}, 
         Scholarpedia {\bf 3(3)}, 3550 (2098).
\bibitem{mackay} R. S.~MacKay, {\it A renormalisation approach to 
         invariant circles in area-preserving maps},
         Physica D {\bf 7}, 283 (1983).
\bibitem{percival} R. S.~MacKay and I. C.~Percival,
         {\it Converse KAM: theory and practice},
         Commun. Math. Phys. {\bf 98}, 469 (1985).
\bibitem{chirikov2000} B. V. Chirikov, {\it Critical perturbation 
          in standard map: a better approximation}, 
            arXiv:nlin/0006021[nlin.CD] (2000).
\bibitem{meiss} J. D.~Meiss, {\it Symplectic maps,
         variational principles, and transport},
         Rev. Mod. Phys. {\bf 64(3)}, 795 (1992).
\bibitem{birth} M. C.~Borja and J.~Haigh,
        {\it The birthday problem},
        Significance {\bf 4(3)}, 124 (2007).


\end{thebibliography}
\end{document}